\documentclass[onecolumn,showpacs,showkeys,preprintnumbers,amsmath,amssymb]{revtex4}

\usepackage{graphicx}
\usepackage{dcolumn}
\usepackage{bm}
\usepackage{epsfig}
\begin{document}

\preprint{IST/CFP 2.2006-M J Pinheiro}

\title[]{Do Maxwell's equations need revision? - A methodological note}
\author{Mario J. Pinheiro}
\address{Department of Physics and Center for Plasma Physics,\&
Instituto Superior Tecnico, Av. Rovisco Pais, \& 1049-001 Lisboa,
Portugal} \email{mpinheiro@ist.utl.pt}





\date{\today}%
\begin{abstract}
We propose a modification of Maxwell's macroscopic fundamental set
of equations in vacuum in order to clarify Faraday's law of
induction. Using this procedure, the Lorentz force is no longer
separate from Maxwell's equations. The Lorentz transformations are
shown to be related to the convective derivative, which is
introduced in the electrodynamics of moving bodies. The new
formulation is in complete agreement with the actual set of
Maxwell's equations for bodies at rest, the only novel feature is
a new kind of electromotive force. Heinrich Hertz was the first to
propose a similar electrodynamic theory of moving bodies, although
its interpretation was based on the existence of the so called
"aether". Examining the problem of a moving circuit with this
procedure, it is shown that a new force of induction should act on
a circuit moving through an inhomogeneous vector potential field.
This overlooked induction force is related to the Aharanov-Bohm
effect, but can also be related to the classical electromagnetic
field when an external magnetic field is acting on the system.
Technological issues, such as the so called Marinov motor are also
addressed.
\end{abstract}
\maketitle
\section{Introduction}

The theory proposed by James Clerk Maxwell successfully unified
optics and the electrodynamics of moving bodies. In 1855 he tried
to unify Faraday's intuitive field lines description and Sir
William Thomson's mathematical analogies to the laws of
hydrodynamics, in particular, making use of his 1842 analogy
relating heat propagation to electrostatic theory. In 1861 Maxwell
proposed a complete set of equations including the displacement
current, from which the electromagnetic wave equation could be
obtained~\cite{Noteh}. Despite its success, however, Maxwell's
equations are still the subject of conceptual difficulties and
controversy. Notwithstanding its widespread technological
applications, it remains particularly painful for scientists and
engineers to apply Michael Faraday's law of
induction~\cite{Crooks,Munley}. Faraday himself gave an empirical
rule to determine when an induced voltage should be expected in a
circuit. The explanation of this rule requires the use of
Maxwell's equations if the magnetic field changes with time, but
the Lorentz force (considered necessary to define the fields
$\mathbf{E}$ and $\mathbf{B}$) if the circuit is
displaced~\cite{Note2}.

The title of this paper is intended to call attention to some of
these conceptual
difficulties~\cite{Mazzoni:01,Graneau:82,Robson:03,Cavalleri:03}.
In our view the changes proposed in this work do not challenge
Maxwell's equations as written for bodies at rest. The novelty of
our approach lies in a new form for the expected electromotive
force and a more systematic presentation of the Maxwell's
fundamental set of equations.

We show that the problem of electromagnetic induction can be more
clearly understood using an appropriate procedure. Although it is
not usually mentioned in the literature, Heinrich Hertz was the
first to propose this procedure and give a systematic treatment of
Maxwell's macroscopic equations for the case of moving bodies.
This inquiry has its origin in the thought that the theory's
current difficulties must be intimately connected to the transport
process of physical quantities.

\section{Electrodynamics of moving bodies}

For a better understanding of Maxwell's fundamental set of
equations we must return to its main experimental sources:
Faraday's law of induction and Amp\`{e}re's law. To facilitate the
analysis we consider all our sources to be in a vacuum. With his
creative mind, Maxwell built his theory on two basic concepts
imported from fluid theory: the notions of circulation and flux.
Whenever the flux of a well-behaved vector field through a given
surface $S$ is calculated, Gauss's theorem is applied. For the
electric field Maxwell's flux equation is (mks units are used
throughout)
\begin{equation}\label{Eq1}
\int_S (\mathbf{E \cdot \mathbf{n}}) dS = \int\int\int_V (\nabla
\cdot \mathbf{E}) dV = \frac{1}{\varepsilon_0} \int\int\int_V \rho
dV.
\end{equation}
From this a differential equation can be obtained, which merely
states the conservation of charge
\begin{equation}\label{Eq2}
\nabla \cdot \mathbf{E} = \frac{\rho}{\varepsilon_0}.
\end{equation}
Applying the same divergence theorem to the flux of the magnetic
field results in another source equation stating the non-existence
of magnetic monopoles):
\begin{equation}\label{Eq3}
\nabla \cdot \mathbf{B} = 0.
\end{equation}
It is usual in textbooks (e.g., Ref~\cite{Jackson}) following the
argument presented by Maxwell himself~\cite{Maxwell}, to assume
that the surface $S$ and its enclosed volume $V$ are fixed to a
stationary frame of reference $\mathbb{R}$. This is the source of
a common misunderstanding in the extension of Maxwell's theory to
moving bodies. Indeed, Maxwell never attempted a systematic
treatment of electromagnetic phenomena in moving
bodies~\cite{Maxwell,Hertz}.

To obtain the other two differential equations, Maxwell turned to
the concept of circulation. From the integral equation we obtain
\begin{equation}\label{Far1}
\oint_{\gamma} (\mathbf{E} \cdot \mathbf{d l}) = - \frac{d
\Phi}{dt},
\end{equation}
where the integration is done along a closed curve $\gamma$ that
borders an open surface $S$. As was the case in the first two
equations, $\gamma$ and $S$ are considered to be at rest in some
frame $\mathbb{R}$, Equation~\ref{Far1} duplicates the results of
Faraday's experiments. In textbooks, Faraday's law of induction is
usually presented in the form of Eq.~\ref{Far1} (e.g.,
Refs.~\cite{Jackson,Landau}). Other authors, however, prefer to
calculate it through the approach $\mathcal{E}=\int \mathbf{F}.d
\mathbf{l}/q$, where $\mathbf{F}$ is the Lorentz force (e.g.,
Ref.~\cite{Crooks} and references therein). By Stoke's theorem,
however, we should have
\begin{equation}\label{Eq5}
\oint_{\gamma} (\mathbf{E'} \cdot \mathbf{d l'}) = \int \int
[\nabla \times \mathbf{E}'] \cdot \mathbf{d S'},
\end{equation}
where the primes denote a well-behaved vector field as measured by
an observer moving along with the circuit, i.e., transported along
the the loop $\gamma$, with speed $v$ relative to frame
$\mathbb{R}$. If the circuit transport is to be considered, the
convective derivative operator, $D/Dt$ should be introduced. Here,
it is expressed in terms of the divergence and curl operators for
greater utility~\cite{Note1}:
\begin{equation}\label{Eq6}
\frac{D}{Dt} \cdot = \frac{\partial}{\partial t} \cdot +
\mathbf{v}(\nabla \cdot) - \nabla \times [ \mathbf{v} \times ]
\cdot
\end{equation}
Applying this operator to Eqs.~\ref{Far1} and ~\ref{Eq5} we obtain
\begin{equation}\label{Eq7}
[\nabla \times \mathbf{E}'] = - \frac{D \mathbf{B}}{D t}= -
\frac{\partial \mathbf{B}}{\partial t} - \nabla \times [\mathbf{B}
\times \mathbf{v}],
\end{equation}
where $\mathbf{E}'$ is the electric field measured by an
instrument attached to the transported frame $\mathbb{R}'$ and
$\mathbf{B}$ is the magnetic field measured in the rest frame. The
analysis of bodies at rest leads us to conclude that time
variation in the magnetic field at a given point are dependent on
the local electric field distribution. Whenever the body is in
motion, however, Heinrich Hertz~\cite{Hertz} explicitly argues the
necessity of following the field along the line of motion and
offers a procedure to carry out this calculation~\cite{Note4}.
Hertz, the discoverer of electric waves, suggested this method to
account correctly for induction phenomena in moving conductors.
Unfortunately, his ideas were not recognized because his theory
did not agree the results of experiments involving displacement of
nonconductors~\cite{Born}.

We can return to the frame $\mathbb{R}$ by transforming the new
electric field according to the rule
\begin{equation}\label{Eq7}
\mathbf{E}' = \mathbf{E} + [\mathbf{v} \times \mathbf{B}],
\end{equation}
which is valid in the non-relativistic limit. In an inertial frame
at rest with respect to two identical charges placed a distance
$r$ apart, each other exerts a force given by $F=q^2/(4 \pi
\varepsilon_0 r^2)$ on the other. This relation serves to {\it
define} the charge $q$ (an axiomatic presentation of Maxwell's
equations is undertaken in Ref.~\cite{Walker}) and is the basis of
electrostatics Coulomb's law when written in the $\mathbb{R}'$
inertial frame still has the form
\begin{equation}\label{Eq8}
\mathbf{F}'=q \mathbf{E}'.
\end{equation}
Now we can return to the $\mathbb{R}$ frame (in practice, the
laboratory frame) in the non-relativistic limit by using
Eq.~\ref{Eq7}. The well-known expression for the Lorentz force
results:
\begin{equation}\label{}
\mathbf{F} = q \mathbf{E}'=q\mathbf{E} + q [ \mathbf{v} \times
\mathbf{B}].
\end{equation}
The Lorentz force is therefore a necessary tool whenever Maxwell's
equations are written for bodies (circuits) at rest. If physical
bodies are not present, the Lorentz force does not need to be
treated as an intrinsic part of Maxwell's equations for a
macroscopic electromagnetic field. In fact, the Lorentz force is a
by-product of Maxwell's equations and the change of variables
required to guarantee their covariance. The usual formulation
presented in textbooks (e.g., Ref.~\cite{Jackson}) fixes the
circuit, $E$ and $B$ to the same reference frame. This is an
inconsistent approach to Faraday's law, since induction from
motion can not be obtained from Maxwell equations.

This relationship (originally proposed by Hertz) nullifies the
widespread argument that when used with the Lorentz force law,
Maxwell's equations describe all electromagnetic phenomena (e.g.,
~\cite{Houser} and references therein).

As we shall gradually come to realize, it is more proper to
rewrite the general form of Maxwell's fourth equations in a given
reference frame as
\begin{equation}\label{Eq10}
[\nabla \times \mathbf{B'}] = \frac{1}{c^2} \frac{D
\mathbf{E}}{Dt} + \mu_0 \mathbf{J},
\end{equation}
in terms of the convective current $\mathbf{J}$. In analogy with
Equation ~\ref{Eq7}, $\mathbf{B}'$ is the magnetic field measured
in the moving frame and $\mathbf{E}$ is the electric field defined
in the rest frame. Expanding the convective derivative we obtain
\begin{equation}\label{Eq12}
c^2 [\nabla \times \mathbf{B'}] = \frac{\partial
\mathbf{E}}{\partial t} + \frac{\rho \mathbf{v}}{\varepsilon_0} -
\nabla \times [\mathbf{v} \times \mathbf{E}] +
\frac{\mathbf{J}}{\varepsilon_0}.
\end{equation}
The first term on the right-hand side is the displacement current
~\cite{Jackson99,Bartlett}, and $\mathbf{J}$ is defined as $\rho
\mathbf{\overline{u}}$ where $\overline{u}$ is the average
velocity of current carriers (conduction electrons) drifting along
the conductor lattice (their typical speed in gold, for example,
is about 1 mm$/$s). One of the benefits  of writing Maxwell's
equation in this form is that there is no longer any need to use
the Lorentz equation to understand the electrodynamics of moving
bodies \cite{Jackson,Maxwell,Lorentz}. Another advantage is that
we recover a higher level of logical consistency in the equations.
The induced electromotive force (emf) is given by
\begin{equation}\label{}
\mathcal{E} = \oint (\mathbf{E}' \cdot d\mathbf{l}')= \int
\int_{S'} [\nabla \times \mathbf{E}'] \cdot d \mathbf{S'}.
\end{equation}
Then, simply by using Eq.~\ref{Eq7} we have
\begin{equation}\label{}
\mathcal{E}=- \int\int \left( \frac{\partial \mathbf{B}}{\partial
t} \cdot d\mathbf{S'} \right) - \int\int_{S'} \nabla \times
[\mathbf{B} \times \mathbf{v}] \cdot d\mathbf{S}'
\end{equation}
and consequently
\begin{equation}\label{}
\mathcal{E} = - \int\int_{S'} \left( \frac{\partial
\mathbf{B}}{\partial t} \cdot d \mathbf{S'} \right) +
\oint_{\gamma} [\mathbf{v} \times \mathbf{B}] \cdot d\mathbf{l}'.
\end{equation}
The induced electromotive force is given as a sum of the
"transformer emf" $\mathcal{E}_{\mathrm flux}$ obtained by
integrating $\partial \mathbf{B}/\partial t$ over the
instantaneous area $S'$ (providing emf even in vacuum, as in
Kersts's betatron)~\cite{Kerst:42}, plus the "motional emf"
$\mathcal{E}_{\mathrm motion}$ obtained with the $[\mathbf{v}
\times \mathbf{B}]$ (a material medium must move through a B-field
to produce an emf). Maxwell's equations thus become complete in
themselves. In order to reconstruct Faraday's induction law, a
similar suggestion was also advanced by Rosen and
Schieber~\cite{Rosen} and Scalon {\it et al}~\cite{Scalon}.

If we redefine the new field as
\begin{equation}\label{}
\mathbf{B}' = \mathbf{B} - [\mathbf{v} \times \mathbf{E}],
\end{equation}
then Maxwell's equation (\ref{Eq12}) can be written as
\begin{equation}\label{}
c^2 [\nabla \times \mathbf{B}] = \frac{\partial
\mathbf{E}}{\partial t} + \frac{\rho
(\mathbf{v}+\mathbf{\overline{u}})}{\varepsilon_0}.
\end{equation}
Notice that the first term on the right-hand side is the
derivative of the local electric field with respect to time. We
can now see why it is imperative to introduce a convective
derivative (which cannot be reduced to the temporal derivative at
a fixed point of space in an inertial frame) into Maxwell's
fundamental equations. Hertz and Lorentz interpreted $v$ as the
velocity relative to the ether, while according to Emile Cohn, a
leading physicist at the University of Strasbourg at the end of
XIX century, $v$ should be interpreted as the velocity of matter
relative to the fixed stars. Even Cohn referred to it as an
absolute velocity (see Ref.~\cite{Darrigol} and references therein
for an historical account).

\section{The convective derivative and the Lorentz transformation}

To round out the picture described above, it must be admitted that
electromagnetism is based on a hybrid foundation; the rigidity of
mechanical laws is interwoven with the notion of flux and
circulation. This is a good reason to come back to the convective
derivative. We would like to note that if we restrain circuit
motion to the $x$-axis, we have
\begin{equation}\label{conv1}
\frac{D }{D t} = \frac{\partial }{\partial t} + v\frac{\partial
}{\partial x}.
\end{equation}
This operator connects the Eulerian (following the moving frame)
and Lagrangian (fixed in a stationary frame) points of view. The
operator $D/Dt$ gives the variation in a given quantity with
respect to time, as measured along the motion. We can denote this
same rate by $D/Dt'$ in the new moving frame, with respect to a
new time coordinate $t'$. By assuming the time dilation effect the
two derivatives can be related by $D/Dt=D/\gamma Dt'$, where
$\gamma=(1-v^2/c^2)^{-1/2}$, and $c$ is the speed of light in
vacuum. It is therefore a simple matter to show that
Eq.~\ref{conv1} and the assumption of linearity in $t$ lead to the
following transformation between $\mathbb{R}$ and
$\mathbb{R}'$)~\cite{Note2}:
\begin{equation}\label{gal1}
t' = \gamma \left( t - \frac{(\mathbf{v} \cdot \mathbf{r})}{c^2}
\right).
\end{equation}
(Substituting Eq.~\ref{gal1} into Eq.~\ref{conv1} recovers an
identity after taking into account time dilation). We therefore
{\it have} to introduce a $\gamma$ factor so that Eq.\ref{gal1}
can verify Eq.~\ref{conv1}, as well to comply with optical
phenomena such as the Michelson-Morley experiment. Using the
convective derivative obliges us to introduce "a suitable change
of variables" (as asserted by Lorentz himself in
Ref.~\cite{Lorentz}). The relation $\gamma D t' = Dt$ implies that
a clock attached to a frame moving with constant speed $v$
relative to a stationary frame will be animated with a slower
rhythm.

By the same token, we can transform the spatial coordinates from
frame $\mathbb{R}$ to frame $\mathbb{R}'$ by applying the
convective derivative, and still obtain an identity if we choose
the appropriate transformation for motion along the $x$-axis:
\begin{equation}\label{Eq19}
x'=\gamma (x-vt).
\end{equation}
Substituting this transformation into Eq.~\ref{conv1}, we have
\begin{equation}\label{Eq20}
\frac{Dx'}{Dt}= \left(\frac{\partial }{\partial t} + v
\frac{\partial }{\partial x} \right)x'.
\end{equation}
Eq.~\ref{Eq20} gives $Dx'/Dt'=0$, as expected. Finally, as
Eq.~\ref{conv1} leads to Eq.~\ref{gal1} (and complies with the
Michelson-Morley experiment), we can see in this procedure a new
derivation of the Lorentz's transformations; the spatial
coordinate transformation is just a consequence of the assumed
symmetry between the inertial frames.

\begin{figure}[htb]
  \includegraphics[scale=0.6]{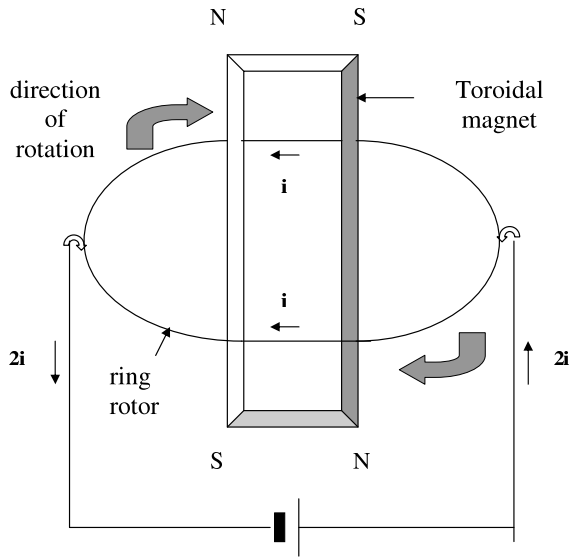}\\
  \caption{Schematic of a Marinov motor.}\label{Fig1}
\end{figure}

Historically, Voldemar Voigt (in 1887) and Sir Joseph Larmor (in
1900)~\cite{Larmor1,Whittaker2} were the first to derive
precursors to the Lorentz transformations (see also
~\cite{Pauli,Kittel}). As the above lines suggest, however, time
as it appears in the Lorentz transformations is a more appropriate
variable than any absolute, "real" variable and is easy to relate
to everyday life experience. This has motivated some proposals to
change to a more "convenient" time variable in discussing and
understanding physical events~\cite{Edwards,Sexl}, and led others
to discuss the twin paradox in the context of an absolute time,
based on cosmological principles~\cite{Roman}. To illustrate the
procedure further, we can relate the charge density in a
"stationary" frame to that in a "moving" frame. Based on the
principle of covariance, we can state that in an inertial frame
$\mathcal{R}'$:
\begin{equation}\label{Eq22}
\nabla \cdot \mathbf{E'} = \frac{\rho'}{\varepsilon_0}.
\end{equation}
By substituting Eq.~\ref{Eq7} into Eq.~\ref{Eq22} above, we obtain
\begin{equation}\label{}
\nabla \cdot \mathbf{E} = \frac{\rho'}{\varepsilon_0} + [\nabla
\times \mathbf{B}] \cdot \mathbf{v}.
\end{equation}
To simplify the problem, we neglect the displacement current and
only consider circuits or particles in translational motion.
Hence, when inserting Eq.~\ref{Eq10} into the above equation we
get
\begin{equation}\label{}
\nabla \cdot \mathbf{E} = \frac{1}{\varepsilon_0} \left( \rho' +
\frac{(\mathbf{v} \cdot \mathbf{J})}{c^2} \right).
\end{equation}
In order to verify the covariance principle we have to modify this
equation as follows:
\begin{equation}\label{Eq17}
\rho' = \left( \rho - \frac{(\mathbf{v} \cdot \mathbf{J})}{c^2}
\right).
\end{equation}
As shown, the above equation is valid in the non-relativistic
limit. Taking into account appropriate scaling factors in the
space and time coordinates (see, for example, the procedure
proposed in Ref.~\cite{Rosen}), the $\gamma$ factor finally
appears in Eq.~\ref{Eq17}.

\section{The null-field force of induction}

To fully explore the consequences of applying the procedure
described above, we must point out that electromotive forces act
on a circuit both in the presence of time-varying magnetic fields
and through motion of the circuit through an inhomogeneous field
according to
\begin{equation}\label{}
\mathbf{E'} = - \nabla \phi - \frac{D}{Dt}\mathbf{A}=-\nabla \phi
- \frac{\partial \mathbf{A}}{\partial t} - (\mathbf{v} \cdot
\mathbf{\nabla}) \mathbf{A}.
\end{equation}
This was already stated by Maxwell (see, for example,
Ref.~\cite{Okun}). Supposing that $v$ is constant, we can use
vector identities to rearrange this into a more easily interpreted
expression:
\begin{equation}\label{Eq27}
\mathbf{E'} = -\nabla \phi -\frac{\partial \mathbf{A}}{\partial t}
+ [\mathbf{v} \times \mathbf{B}] -\nabla (\mathbf{v} \cdot
\mathbf{A}).
\end{equation}
The gradient of a scalar field $\phi$ is a conservative vector
field, so the resulting electromotive force receives contributions
from three terms. Aside from the well-known forces of induction
due to a time-varying, local magnetic field and the motional emf
(the second and third terms of Eq.~\ref{Eq27}), there exists a
third force of induction given by
\begin{equation}\label{Eq28}
\mathbf{F} = -q \nabla (\mathbf{v} \cdot \mathbf{A})
\end{equation}
This term, which we might call the null field force of induction,
demonstrates the interesting possibility of generating a force of
induction even in the absence of a magnetic field provided that a
position-dependent vector potential $A$ can be produced.

\begin{figure}[htb]
  \includegraphics[scale=0.7]{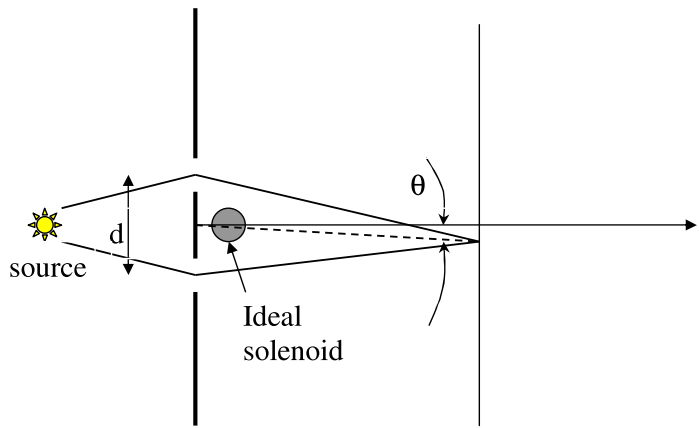}\\
  \caption{Scheme of the electron beam deflection on the double-slit experiment. An infinite solenoid
  is placed between the two slits and creates on the
  exterior a vector potential $\mathbf{A}$. }\label{Fig2}
\end{figure}

Although still controversial, an induction motor based on the
action of null field induction was apparently conceived by S.
Marinov~\cite{Marinov}. The concept is simple (see
Fig.~\ref{Fig1}): a permanent magnet of toroidal shape (enclosing
its own B-field) is placed in a fixed position inside a conducting
ring. The ring is supported by bearings and a direct current is
provided through sliding contacts on opposite sides, driving the
ring in continuous motion.

An advanced explanation for the forces driving this motor can be
based on the this third force of induction~\cite{Wesley}. The
existence of an applied torque in the Marinov motor apparently
points out that the Lorentz force law is not enough to describe
all observable electromagnetic forces~\cite{Phipps,Onoochin}. This
interesting and unusual property is also seen in the Aharonov-Bohm
effect~\cite{Bohm}. Aharonov and Bohm devised a hypothetical
diffraction grating experiment, in which the vector potential
would influence not only the interference pattern but also the
momentum of the diffracted beam~\cite{Wesley,Bohm} (see
Fig.~\ref{Fig2}). It is worthwhile to note that there exist two
different types of effects which are not always clearly
distinguished in the literature: i) a deflection of the entire
interference pattern due to a classical force; and ii) a
deflection of the double-slit interference pattern
only~\cite{Boyer1}. In the two-slit experiment considered by
Aharonov and Bohm~\cite{Bohm}, an infinite solenoid is placed
between the two slits so that the magnetic field is zero on all
paths. If we consider the action of the null field induction force
on the electron beam we obtain
\begin{equation}\label{}
F = q E_3 = -q \int \frac{1}{r} \frac{\partial }{\partial
\theta}(v A_{\theta}) r \frac{d \theta}{\Delta t}=i[(vA)_{\theta +
\delta \theta} - (vA)_{\theta}],
\end{equation}
where $E_3$ is the null field emf contribution previously referred
to, $i$ is the electric current, and $\theta$ is the deflection
angle of the electrons (see Ref.~\cite{Boyer1} for a more detailed
explanation). We have applied an appropriate change of variables
$dq \mathbf{v}=i d\mathbf{s}$, where $d\mathbf{s}$ is the circuit
line element. It is worthwhile to mention that this term can also
generate a classical force if $A=Bx$ (in the presence of a
magnetic field), involving a deflection of the electron beam.
Without an external field ($\mathbf{B}=0$), on the other hand, it
can generate the Aharonov-Bohm effect which involves no average
deflection of the electron beam, but only a deflection of the
double-slit pattern ~\cite{Boyer1}. It is clear now that this
electromotive force works in open currents, but will it work in
closed loops? According to the properties of the gradient
operator, we should also expect a null field electromotive force
in loop currents. Unfortunately, there is still no clear
experimental evidence for this motive force in closed loops.
Nevertheless, Trammel's~\cite{Trammel} discussion of the
Aharonov-Bohm paradox and referral to the peculiar quantum
significance of the vector potential is not quite true. Even in
the classical limit, this same paradoxical behavior occurs. This
point is also clearly discussed in Ref.~\cite{Boyer1}. In
particular, Trammel shows that the forces between a charged
particle $q$, $\mathbf{F}_q$, and a current-carrying body,
$\mathbf{F}_j$, are not equal and opposite but obey the relation
(see also note~\cite{Note30a} and Ref.~\cite{Calkin})
\begin{equation}\label{Tram1}
\mathbf{F}_q + \mathbf{F}_j = - \frac{D}{Dt}q \mathbf{A}.
\end{equation}
As the interaction between charges and loops must be
counterbalanced by mechanical means, the corresponding mechanical
force should be
\begin{equation}\label{}
\mathbf{F}_{ME}=-\mathbf{F_q}-\mathbf{F_j}=\frac{D}{Dt}(q\mathbf{A}).
\end{equation}
Applying Eq.~\ref{Tram1} to determine the force that a charged
particle exerts on a structural ring allows one to subsequently
obtain the angular velocity of the transmitted
rotation~\cite{Trammel}.

\section{Further considerations}

As a matter of fact, there is a group of affine transformations
which leave Maxwell's equations invariant provided appropriate
definitions of the magnetic and electric fields are
adopted~\cite{Miller}. Maxwell's equations, while containing all
electrodynamics, are an indeterminate system. Usually, the
following expressions are used for the basic fields $\mathbf{E}$
and $\mathbf{B}$ in an isotropic medium (and inertial frame):
\begin{equation}\label{}
\begin{array}{ccc}
  \mathbf{D}=\varepsilon \mathbf{E}; & \mathbf{B}=\mu \mathbf{H}; & \mathbf{J}=\sigma (\mathbf{E}+\mathbf{E}^{ext}), \\
\end{array}
\end{equation}
where the quantities $\varepsilon$ and $\mu$ characterize the
properties of the medium and $\mathbf{E}^{ext}$ denotes an
external electric field. The dependencies
$\mathbf{D}=\mathbf{D}(\mathbf{E})$ and
$\mathbf{H}=\mathbf{H}(\mathbf{B})$ result from the Poynting
theorem. Some other more general relation such as
$\mathbf{D}=\mathbf{D}(\mathbf{E},\mathbf{B})$ or
$\mathbf{H}=\mathbf{H}(\mathbf{E},\mathbf{B})$, however, is also
possible~\cite{Paris} .

These constitutive equations must be listed among Maxwell's
fundamental equations. These relations do not, however, ensure
that the three-dimensional vectors of the electric and magnetic
fields, $\mathbf{E}$ and $\mathbf{B}$, will obey the correct
relativistic transformations, only the 4-vectors $E^{\alpha}$ and
$B^{\alpha}$ are well-defined quantities. This question was well
treated in Ref.~\cite{Ivezic}. It is still unclear how one should
extend relativistic electrodynamics to material media, and in
particular there are recognized difficulties when the medium is in
rotation~\cite{Pelegrinni}.

It has been argued historically that relativity emerged because
Maxwell's equations are not invariant under Galilean
transformations, but this statement must be treated with due
caution~\cite{Miller,Goldin,Gomberoff}. According to Einstein's
theory of relativity, the operational meaning of time is
essentially that of a system of local clocks synchronized with
each other by optical means. Of course, to synchronize these
clocks some procedure must be specified: Einstein assumed
communication through light, with equal speeds in both directions.
Other authors, not satisfied with this state of affairs, have
proposed alternative theories~\cite{Edwards,Sexl} in which time
and space coordinates on a "moving" frame are {\it defined} by the
"stationary" observer in frame $\mathcal{R}$ through optical
means.

We are far from the clear consensus on the local meaning of time
envisioned by Eddington: the truth is that the definition of time
is a problem of extreme complexity~\cite{Eddington}.

\section{Conclusion}

Through an appropriate modification of Maxwell's equations using
the convective derivative Faraday's law of induction is clarified,
giving new insight into electromagnetic phenomena. The convective
derivative operator can serve as a foundation for the Lorentz
transformations, and the difficulty of relating the time variable
to daily experience in a clear-cut manner is intrinsically related
to these issues.

It was shown that a third force of induction, the null-field
force, must act on an electric circuit in the presence of a vector
potential. This interesting phenomenon is likely to be useful in
numerous practical applications.

\begin{acknowledgments}This work was supported primarily by the
Rectorate of the Technical University of Lisbon and the
Funda\c{c}\~{a}o Calouste Gulbenkian.
\end{acknowledgments}

\bibliographystyle{amsplain}
\bibliography{Doc2}
\end{document}